\documentclass[
]{ceurart}

\usepackage{listings}
\usepackage{ulem}
\usepackage{tikz}
\usetikzlibrary{matrix,decorations,arrows,shapes,automata,backgrounds,fit,calc,petri,patterns,positioning}

\usepackage{graphicx}
\usepackage{xspace}
\usepackage{tikz-network}
\usepackage{tabu}
\usepackage{amsmath}

\newcommand{\Nat}{\mathbb{I\!\!N}}

\newcommand{\nsymbol}{\mathbb{N}}

\newcommand{\abbrev}[1]{#1, \relax}
\newcommand{\ie}[0]{\abbrev{\textit{i.e.}}}

\newcommand{\tuple}[1]{\langle #1 \rangle}
\newcommand{\Bool}{{\mathbb B}}
 
\newcommand{\evo}[0]{\longrightarrow}
\newcommand{\xevo}[1]{\stackrel{#1}{\evo}}
\newcommand{\arc}{\ensuremath{\,\tikz[baseline=-0.5ex,scale=0.65,thin]{\draw[->,>=open triangle 45](0,0)to[bend left=0] (1,0);}\,}}
\newcommand{\shortarc}{\ensuremath{\,\tikz[baseline=-0.5ex,scale=0.45,thin]{\draw[->,>=open triangle 45](0,0)to[bend left=0] (1,0);}\,}}
\newcommand{\eqdef}{\overset{\text{def}}{=\mathrel{\mkern-3mu}=}}

\begin{document}

\copyrightyear{2025}
\copyrightclause{Copyright for this paper by its authors.  Use permitted under Creative Commons License Attribution 4.0  International (CC BY 4.0).}
\conference{PNSE'25, International Workshop on Petri Nets and Software Engineering, 2025}
\setcounter{page}{1234}
\pagestyle{headings} 


\title{Glucagon and insulin production in pancreatic cells modelled using Petri nets and Boolean networks.}

\author[1]{Kamila Barylska}[%
]

\author[2]{Franck Delaplace}[%
]

\author[1]{Anna Gogoli{\'n}ska}[%
email=leii@mat.umk.pl,
]

\author[3]{Ewa Pa{\'n}kowska}[%
]

\address[1]{Nicolaus Copernicus University in Toru{\'n}}
\address[2]{Paris-Saclay University - University Evry}
\address[3]{Institute of Diabetology, Warsaw}

\begin{abstract}
Diabetes is a civilization chronic disease characterized by a constant elevated concentration of glucose in the blood. Many processes are involved in the glucose regulation, and their interactions are very complex. To better understand those processes we set ourselves a goal to create a Petri net model of the glucose regulation in the whole body. So far we have managed to create a model of glycolysis and synthesis of glucose in the liver, and the general overview models of the glucose regulation in a healthy and diabetic person.\\
In this paper we introduce Petri nets models of insulin secretion in $\beta$ cell of the pancreas, and glucagon in the pancreas $\alpha$ cells. Those two hormones have mutually opposite effects: insulin preventing hyperglycemia, and glucagon preventing hypoglycemia. Understanding the mechanisms of insulin and glucagon secretion constitutes the basis for understanding diabetes. We also present a model in which both processes occur together, depending on the blood glucose level. The dynamics of each model is analysed.
Additionally, we transform the overall insulin and glucagon secretion system to a Boolean network, following standard transformation rules.
\end{abstract}


\maketitle
\vspace*{-0.2cm}
\section{Introduction}

Maintaining energy balance is a fundamental goal of every living organism. Energy requirements are determined by many factors: body size, level of physical activity, age.
It also depends of external conditions in which the body functions require constant adaptation to changes. These energy-consuming processes, e.g., maintaining a constant body temperature when the ambient temperature drops, require continuous access to energy material, which is glucose.

Energy is derived from the oxidation of carbohydrates, fats, and proteins. At the cellular level, glucose is the primary fuel for energy production. Therefore, its concentration in the bloodstream is maintained within a specific range, not lower than 70 mg/dl and not higher than 140 mg/dl. It represents a constant balance between glucose entering the blood, primarily from the gut after meals, and from the liver, as well as glucose uptake by peripheral tissues. The central nervous system continuously consumes up to 60\% of the glucose resources in the blood, so the blood glucose level is tightly regulated to ensure adequate glucose delivery to the brain. Peripheral tissues,  especially muscles, take up 50\% of an oral glucose load (after meals), and at the same time, the glucose is released by the liver. This fuel and energy homeostasis is regulated by the neuroendocrine system. This process occurs on various levels, where the central and autonomic nervous systems play a dominant, integrative role.

Disruptions in the metabolic balance of the body regarding glucose metabolism regulation generate many diseases. When the blood glucose level is too high, increasing causes the development of diseases, one of which is diabetes. On the other hand, when blood glucose levels are too low, can disrupt the functioning of the central nervous system (CNS). During prolonged hypoglycemia (below 40 mg/dl), significant impairments in consciousness and damage to CNS cells may occur. 

Diabetes is a group of different diseases for which the common denominator is hyperglycemia (excess blood glucose). Diabetes is one of civilization's diseases. In diabetes, hyperglycemia occurs through various mechanisms. This happens on different levels: impaired receptor response to insulin (insulin resistance, which often arises from overnutrition and obesity), overproduction of hyperglycemic hormones, prolonged stimulation of the sympathetic nervous system, for example, during chronic stress. In type 1 diabetes, due to an autoimmune process (inflammatory state affecting beta cells of the pancreas), apoptosis of beta cells occurs, and the ability to produce and secrete insulin is irreversibly lost. 
The basic role in the regulation of glycemia is played by insulin, which is the only hormone that lowers blood glucose levels. The other hormone - glucagon plays an essential role in hepatic glucose output. In this way, it increases blood glucose levels.
Together, insulin and glucagon work in tandem to maintain homeostasis in glucose metabolism, ensuring that the body has a stable energy supply while preventing the detrimental effects of excessive or insufficient blood sugar levels. Understanding the dynamics of insulin secretion and the role of $\beta$-cells and $\alpha$-cells is essential for developing strategies to address conditions such as diabetes and metabolic syndrome.

The following work presents the mathematical model of the secretion of two opposite hormones responsible for keeping the glycemia level in the normal range: insulin and glucagon. This is the first attempt by the Petri Net model to describe the secretion hormone process at the cellular level, which is one of the links in the entire metabolic balance regulation process.

Our long-term goal is to create a simple and intuitive mathematical model representing all the processes taking place in the body of a healthy person.
This model should be easily analysable and clear, but at the same time capable of representing complex activities consisting of interactions between many components. 
In our opinion, Petri nets (PNs) constitute an appropriate tool for this purpose. Due to PNs intuitive graphical representation and mathematical properties, the model could be easy utilized by people without mathematical background, for instance biologists, doctors, diabetes educators or~patients.
This allow for a better understanding of the processes occurring in a human body, and predicting new therapeutic targets and designing drug therapies by in-depth analysis and simulations.
We are aware, that our goal is ambitious and would not be reached at once.

Our initial efforts to achieve the desired result have been described in the following papers:~\cite{BG1}, where a Petri net modelling the process of glycolysis and glucose synthesis in the liver was presented, and ~\cite{BG2}, which shows the basic models of glycemia control processes occurring in the bodies of healthy and diabetic people.

The next part of our work are Petri nets models of insulin secretion in $\beta$-cell of the pancreas, and glucagon in $\alpha$-cells. Moreover, we combined both models and performed a detailed analysis of the resulting comprehensive model. Additionally, we show a Boolean network that models exactly the same phenomena. 
After the introduction, we recall the basic concepts of Petri nets. The following section contains the biological basis, while in the subsequent sections we introduce models of insulin secretion, glucagon secrection, and the combined model. In Section~\ref{bn} we present a Boolean network obtained from the overall model with the use of logic formulas based on logical formulas corresponding to the semantics of Petri nets.  
The paper ends with a summary and future plans.

\section{Preliminaries}
\label{s.prel}

Let us begin with recalling basic mathematical concepts, which are important
 to understand the paper.
\subsection{Petri nets}
\label{s.prel.pn}
In this section we recall some  concepts and definitions used  throughout this paper, as well as the basic notions concerning Petri nets and its properties~\cite{primer,petri1,Murata,Reisig,petri2,trap}.

The set of non-negative integers is denoted by $\Nat$.
Given a set X, the cardinality (number of elements) of $X$ is denoted by $|X|$.
A function $\mu : X \to \Nat$  may also be considered as a vector in $\Nat^{|X|}$.  

A {\it finite labelled transition system} with initial state is a tuple $TS=(S,\to,T,s_0)$ with:
\begin{itemize}
\item nodes $S$ (a~finite set of {\it states}),
\item edge {\it labels} $T$ (a~finite set of {\it letters}),
\item {\it edges} $\to\,\subseteq(S\times T\times S)$,
\item an {\it initial state} $s_0\in S$.
\end{itemize}
A label $t$ is {\it enabled} at $s\in S$, denoted by $s[t\rangle$, if $\exists s'\in S\colon(s,t,s')\in\,\to$.
A state $s'$ is {\it reachable} from~$s$ through the execution of $\sigma\in T^*$,
denoted by $s[\sigma\rangle s'$, if there is a directed path from~$s$ to~$s'$ whose edges are labelled consecutively by $\sigma$\footnote{Recall that 
a sequence $(e_0,e_1,\ldots,e_{n-1})$ of edges is called a~{\it(directed) path} 
if there exist nodes
$v_0,v_1,\ldots,v_n\in S$ such that $\forall_{k\in \{0,1,\ldots,n-1\}}$ we have 
$(v_k,e_k,v_{k+1})\in \to$.}.
The set of states reachable from $s$ is denoted by $[s\rangle$.
A sequence $\sigma\in T^*$ is {\it allowed}, or {\it firable}, from a~state $s$,
denoted by $s[\sigma\rangle$, if there is some state $s'$ such that $s[\sigma\rangle s'$.

An {\it (initially marked) Petri net} (PN) is denoted as $N=(P,T,F,M_0)$ where:
\begin{itemize}
\item  $P$ is a finite set of places,
\item $T$ is a finite set of transitions,
\item $F$ is the flow function $F\colon((P\times T)\cup(T\times P))\to\nsymbol$ 
specifying the arc weights,
\item $M_0$ is the initial marking
(where a marking is a mapping $M\colon P\to\nsymbol$, indicating the number of tokens in each place).
\end{itemize}

A transition $t\in T$ is {\it enabled} at a marking $M$,
denoted by $M[t\rangle$, if $\forall p\in P\colon M(p)\geq F(p,t)$.
The {\it execution} ({\it firing}) of $t$ leads from $M$ to $M'$, denoted by $M[t\rangle M'$,
if $M[t\rangle$ and $M'(p)=M(p)-F(p,t)+F(t,p)$.
This can be extended, as usual, to $M[\sigma\rangle M'$ for sequences $\sigma\in T^*$,
and $[M\rangle$ denotes the set of markings {\it reachable} from $M$.
We call a marking $M$ {\it deadlock} if it does not enable any transition, i.e. for every $t\in T$ we have $\exists p\in P\colon M(p)< F(p,t)$.


Let us now recall some basic properties of Petri nets. 
A Petri net $N=(P,T,F,M_0)$ is called: 
\begin{itemize}
\item {\bf $k$-bounded} for some $k$ if {$\forall_{M\in[M_0\rangle} \forall_{p\in P}\; M(p)\leq k$},
\item {\bf bounded} if {$\exists_{k\in\Nat} \forall_{M\in[M_0\rangle}
\forall_{p\in P}\; M(p)\leq k$ } (\ie there exists a natural number such that, for all reachable markings, the number of tokens in each place does not exceed that number, 
which allows to say that $[M_0\rangle$ is finite),
\item {\bf safe} if it is  $1$-bounded,
\item {\bf weakly live} if { $\forall_{t\in T} \exists_{M\in [M_0\rangle} \;M[t\rangle$} 
(every transition is reachable from the initial state),
\item {\bf live} $\forall_{t\in T}\forall_{M\in[M_0\rangle} \exists_{M'\in[M\rangle} \; M[t\rangle$ (no transitions can be made unfireable),
\item {\bf reversible} $\forall_{M\in [M_0\rangle}\; M_0\in [M\rangle $
($M_0$ always remains reachable).
\end{itemize}

If $x\in P\cup T$, the {\it pre-set} ${}^\bullet x$ and {\it post-set} $x^\bullet$ of $x$ are defined as: 
\begin{itemize}
\item [] ${}^\bullet x=\{y\in T\cup P\mid F(y,x)>0\}$,
$x^\bullet=\{y\in T\cup P\mid F(x,y)>0\}$.
\end{itemize}
We extend the above notations to sets as follows: for $S \subseteq P \cup T$:
\begin{itemize}
\item [] $^\bullet S = \bigcup_{x \in S} {^\bullet x}\;\;\;$ and
$\;\;\;S^\bullet = \bigcup_{x \in S} x^\bullet$.
\end{itemize}  
Having $x,y\in P\cup T$, if $F(x,y)>0$, we say that $x$ is an {\it input place} to $y$ if~$x\in P$, and an {\it input transition} to $y$ if $x\in T$. In that case also, $y$ is an {\it output transition} from $x$ if $y\in T$, and an {\it output place} from $x$ if~$y\in P$.
A {\it self-loop} in a Petri net is when a place is both an input and output place of a transition.

Let $N = (P, T, F, M_0)$ be a Petri net. 
Assume that $P=\{p_1,\ldots,p_n\}$ and $T=\{t_1,\ldots,t_m\}$. Then:
\begin{itemize}
\item {\it input matrix} for $N$ is a matrix $C^-=(a_{i,j})_{n\times m}$, 
where $a_{i,j}=F(p_i,t_j)$,
\item {\it input matrix} for $N$ is a matrix $C^+=(a_{i,j})_{n\times m}$, 
where $a_{i,j}=F(t_j,p_i)$,
\item {\it incidence matrix} for $N$ is a matrix $C=C^+-C^-$, 
\end{itemize}
A $T$-{\it invariant} is a vector $x\in \Nat_m$, such that $C\cdot x=0$, which indicates a possible loop in the net, \ie a~sequence of transitions whose net effect is null (\ie which leads back to the marking it starts in).

\begin{figure}[htb]
\begin{center}
\hbox{}
\begin{tikzpicture}
\node[draw,minimum size=0.5cm](a)at(0,1.5){$a$};
\node[draw,minimum size=0.5cm](b)at(2,1.5){$b$};
\node[draw,minimum size=0.5cm](c)at(3,1.5){$c$};
\node[circle,draw,thick,minimum size=0.4cm](s0)at(1,3){};\filldraw[black](1,3)circle(2pt);\draw(0.5,3)node{$p$};
\node[circle,draw,minimum size=0.4cm](s1)at(1,2){};\filldraw[black](1,2)circle(2pt);
\node[circle,draw,minimum size=0.4cm](s2)at(1,1){};
\node[circle,draw,minimum size=0.4cm](s3)at(1,0){};
\draw[-latex](s0)--(a);
\draw[-latex](s0)--(b);
\draw[-latex](s1)--(a);
\draw[-latex](b)--(s1);
\draw[-latex](a)--(s2);
\draw[-latex](s2)--(b);
\draw[-latex](a)--(s3);
\draw[-latex](b)--(s3);
\draw[-latex](s3)to[out=0,in=270](c);
\draw[-latex](c)to[out=90,in=0](s0);
\end{tikzpicture}\hspace*{1cm}
\raisebox{1cm}{\begin{tikzpicture}[scale=0.6]
\node[circle,fill=black!100,inner sep=0.05cm](0)at(0,2)[label=above left:$M_0$]{};
\node[circle,fill=black!100,inner sep=0.05cm](1)at(2,2)[label=above:]{};
\node[circle,fill=black!100,inner sep=0.05cm](2)at(2,0)[label=above:]{};
\node[circle,fill=black!100,inner sep=0.05cm](3)at(0,0)[label=above:]{};
\draw[-triangle 45](0)to[]node[above,swap,inner sep=2pt]{$a$}(1);
\draw[-triangle 45](1)to[]node[right,swap,inner sep=2pt]{$c$}(2);
\draw[-triangle 45](2)to[]node[above,swap,inner sep=2pt]{$b$}(3);
\draw[-triangle 45](3)to[]node[left,swap,inner sep=2pt]{$c$}(0);
\end{tikzpicture}}
\caption{A Petri net (lhs) and its reachability graph (rhs).}\label{netgraph}
\end{center}
\end{figure}
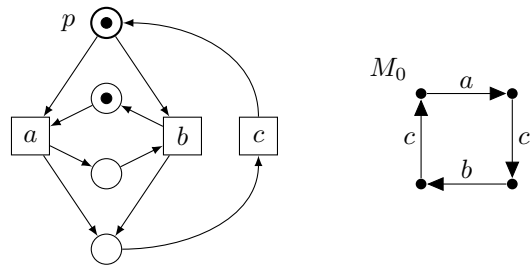

The {\it reachability graph} $RG(N)$ of a bounded Petri net $N$ is the finite labelled transition system with:
\begin{itemize}
\item set of nodes $[M_0\rangle$,
\item label set $T$,
\item set of edges $\{(M,t,M')\mid M,M'\in[M_0\rangle\land M[t\rangle M'\}$,
\item initial state $M_0$.
\end{itemize}
Figure~\ref{netgraph} depicts an example of a Petri net and its reachability graph.

\vspace*{-0.5cm}
Note that the reachability graph of a bounded Petri net captures the exact information about the reachable markings of the net, and therefore reflects the entire behaviour of the given net. 
Figure~\ref{netgraph} depicts a Petri net, together with its reachability graph. 
Reachability graphs of real biological systems are usually quite large and therefore difficult to analyse. 
To deal with this inconvenience, we use reduced reachability graphs, called {\it stubborn reduced reachability graphs}, created on the basis of partial order reduction techniques, where not all interleaving sequences of
concurrent behaviour are considered. 
As~a~result of the reduction only a subset of the complete reachability graph is constructed, nevertheless it still allows the discussion of certain properties, in particular: it preserves all deadlock states and the whole cyclic behaviour.
 
The reduction of a reachability graph to
 a stubborn reduced reachability graph proceeds as follows:
\begin{enumerate}
\item For a given marking, determine a set of "independent" transitions (called {\it stubborn set}), such that their behaviour cannot be influenced by any transitions from the complement 
of this set (\ie~\textit{excluded} transitions). 
Additionally, the following conditions must hold: any sequence of excluded transitions cannot enable or disable an included transition (hence their firing can be postponed) and the set contains at least one enabled transition.
\item Compute a \textit{stubborn reduced reachability graph}, using a variation of a standard algorithm: 
at~each marking (node), instead of firing all enabled transitions, 
 only transitions of a stubborn set are fired.
\end{enumerate}
The notion of stubborn sets capture the lack of interaction between transitions,  and such excluded transitions
 may not be interesting from our point of view (for instance in case of  biological systems). Executions of transitions from outside a stubborn set could be postponed because it does not affect the merits of the system's behaviour\footnote{We do not provide detailed definitions and 
properties here, interested readers are referred to the literature (a.o.: \cite{valmari1}, \cite{valmari2}, \cite{heiner}, \cite{charlie}).}.

\subsection{Boolean networks}
\label{s.prel.bn}
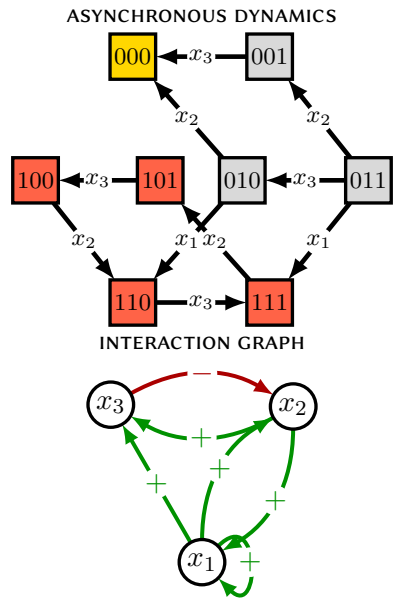
\begin{figure}[h]
	\begin{tabu}{X[c,m]}
		\textsc{asynchronous dynamics}\\
		\begin{tikzpicture}[scale=0.8]
			\definecolor{color1}{HTML}{FFD700} 
			\definecolor{color2}{gray}{0.85}
			\definecolor{color3}{HTML}{FF6347} 
			\definecolor{color4}{gray}{0.}
			\SetVertexStyle[FillColor=white,MinSize=0.6cm,LineWidth=1.25pt, Shape=rectangle, TextFont=\footnotesize]
			\SetEdgeStyle[TextFont=\footnotesize,LineWidth=1.75] \SetDistanceScale{2.5}
			\Vertex[x=0.64, y=1.63, color=color1, Math=true, label=000]{0}
			\Vertex[x=1.54, y=1.63, color=color2, Math=true, label=001]{1}
			\Vertex[x=1.36, y=0.81, color=color2, Math=true, label=010]{2}
			\Vertex[x=0.64, y=0, color=color3, Math=true, label=110]{6}
			\Vertex[x=2.19, y=0.81, color=color2, Math=true, label=011]{3}
			\Vertex[x=1.54, y=0, color=color3, Math=true, label=111]{7}
			\Vertex[x=0, y=0.81, color=color3, Math=true, label=100]{4}
			\Vertex[x=0.82, y=0.81, color=color3, Math=true, label=101]{5}
			\Edge[label=x_1, color=color4, bend=0, Direct=true, Math=true](2)(6)
			\Edge[label=x_3, color=color4, bend=0, Direct=true, Math=true](3)(2)
			\Edge[label=x_1, color=color4, bend=0, Direct=true, Math=true](3)(7)
			\Edge[label=x_3, color=color4, bend=0, Direct=true, Math=true](6)(7)
			\Edge[label=x_2, color=color4, bend=0, Direct=true, Math=true](4)(6)
			\Edge[label=x_3, color=color4, bend=0, Direct=true, Math=true](5)(4)
			\Edge[label=x_2, color=color4, bend=0, Direct=true, Math=true](7)(5)
			\Edge[label=x_2, color=color4, bend=0, Direct=true, Math=true](3)(1)
			\Edge[label=x_3, color=color4, bend=0, Direct=true, Math=true](1)(0)
			\Edge[label=x_2, color=color4, bend=0, Direct=true, Math=true](2)(0)
		\end{tikzpicture}\\
		\textsc{interaction graph}
		\\
		\begin{tikzpicture}[scale=0.8]
			\definecolor{color1}{gray}{1.}
			\definecolor{color2}{rgb}{0., 0.57, 0.}
			\definecolor{color3}{rgb}{0.666667, 0., 0.}
			\SetVertexStyle[FillColor=white,MinSize=0.6cm,LineWidth=1.25pt, Shape=circle, TextFont=\normalsize]
			\SetEdgeStyle[TextFont=\normalsize,LineWidth=1.5] \SetDistanceScale{3.}
			\Vertex[x=0.50, y=0, color=color1, Math=true, label=x_1]{x1}
			\Vertex[x=1.00, y=0.86, color=color1, Math=true, label=x_2]{x2}
			\Vertex[x=0, y=0.87, color=color1, Math=true, label=x_3]{x3}
			\Edge[label=+, color=color2, bend=30, Direct=true, Math=true](x1)(x1)
			\Edge[label=+, color=color2, bend=30, Direct=true, Math=true](x1)(x2)
			\Edge[label=+, color=color2, bend=0, Direct=true, Math=true](x1)(x3)
			\Edge[label=+, color=color2, bend=30, Direct=true, Math=true](x2)(x1)
			\Edge[label=+, color=color2, bend=30, Direct=true, Math=true](x2)(x3)
			\Edge[label=-, color=color3, bend=30, Direct=true, Math=true](x3)(x2)
		\end{tikzpicture}
	\end{tabu}
	\caption{Asynchronous dynamics and equilibria (in different colors for each attractor), interaction graph}
	\label{fig:boon1}
\end{figure}

Boolean Network study applied to biological systems studies was pioneered by Stuart Kaufman~\citep{kauffman1969,glass1973} and Ren\'e Thomas~\citep{thomas1973} in 70s.
The objective was to develop a formal modeling framework for studying the dynamics of genetic networks. Currently, this modelling framework is regarded as a gold standard for studying biological systems. Formally, a {\it Boolean network} is a dynamical discrete system operating on Boolean variables $X$ and defined as a system of Boolean equations of the form: $ x_i = {f_{i}}(x_1,\ldots,x_n)$, $1 \leq i \leq n$ where each $f_i$ is a logical propositional formula. We present in Figure~\ref{fig:boon1} the dynamics of the following Boolean network:
\begin{equation}
	F= \left\{
	\begin{aligned}
		x_1 &=\left(x_1 \lor x_2\right) \\
		x_2 &=\left(\neg x_3\land x_1\right)\\
		x_3 &=\left(x_2\land x_1\right)
	\end{aligned}\right.
	\label{eq:boon}
\end{equation}
A Boolean {\it state} of $s$ is an interpretation of the variables of $X$ in Boolean i.e. $s:X \to \Bool, \Bool = \{0,1\}$, and $S_X = ( X \to \mathbb{B})$ is  the set of all states.

The analysis of a Boolean network is primarily focused on the dynamics to investigate its behaviour.  The Boolean dynamics provide the full description of all effective trajectories, thereby ensuring a~comprehensive and thorough investigation of all possible dynamical scenarios.

Formally, a {\it model of dynamics} is defined by a labelled transition system where for each transition the states of agents are updated according to an updating policy defined by a {\it mode}. Two modes are frequently used in modelling: the {\it asynchronous} mode where one variable exactly is updated per transition and the {\it synchronous} mode where all the variables are updated per transition. As example, the {\it asynchronous} mode leads to the following labelled transition system 
(with no initial state specified) $(S_X,\evo,X)$ 
where the updated agent $\evo \subseteq S_X \times X \times S_X$ labels the transition relation, $\xevo{x_i}$ such that:
$$s_1 \xevo{x_i} s_2 \eqdef s_1\neq s_2 \land s_2(x_i)=f_i(s_1) \land \forall x_j \in X \setminus \{x_i\}: s_2(x_j)=s_1(x_j).$$
A key objective of the analysis is to assess the equilibrium of the modelled network to gain insight into its long-term behaviour.
Basically, a state $s$ is an {\it equilibrium} for $\evo$, if it can be infinitely reached once met, i.e.  $\forall s'\in S_X:s \evo^* s' \implies s' \evo^* s.$ The aim is to identify and characterize the attractors, which are defined as sets of mutually reachable equilibria.

The attractors corresponding to a single state are called {\it stable states} and can be efficiently and quickly computed by symbolic method while the computation of the other attractors (cyclic) remains exponential in general, limiting their investigation to small networks.

Finally, an {\it interaction graph} $\tuple{X,\arc}$ portrays the causal interactions between variables of a Boolean network.
An interaction $x_i \shortarc x_j$ exists if and only if $x_i$ occurs as literal in a minimal Disjunctive Normal Form (\textsc{dnf}) of $f_j$, i.e. $x_i \shortarc x_j \eqdef x_i \in V(\textsc{dnf}(f_j))$. The interactions are refined by a sign stipulating the nature of the interaction: positive ($+$) or negative ($-$). Figure~\ref{fig:boon1} illustrates the asynchronous dynamics of Boolean network~\eqref{eq:boon} leading to a cyclic attractor (red) and a stable state (yellow), and its signed interaction graph.

In this study, the Boolean network is systematically derived from the Petri net following standard transformation rules. This alternative formalism, which maintains semantic consistency with the original Petri net, facilitates the systematic characterization of stable states interpreted as steady molecular states, each potentially associated with distinct phenotypic outcomes.
%
%
%
%
%
%

\section{Biological basis}
\label{bb}
\begin{figure}[htb]
\begin{center}
\includegraphics[scale=0.53]{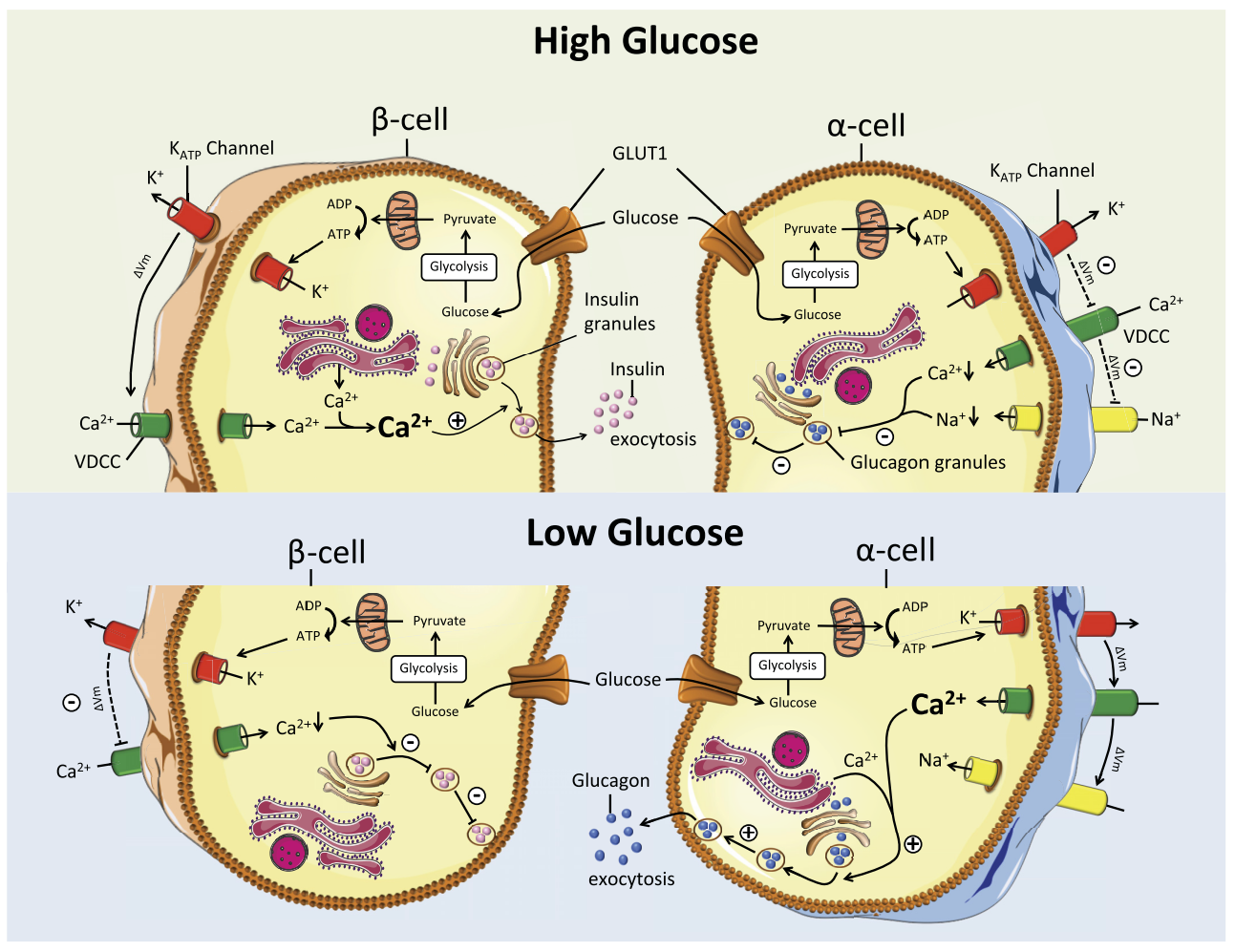}
\label{f.sch}
\end{center}
\vspace*{-0.3cm}
\caption{Schematic on the glucose-dependent regulation of glucagon and insulin secretion~\cite{bb1}.}
\label{f.mi}
\end{figure}

In the following subsections the mechanisms leading to insulin secretion in $\beta$-cells and glucagon secretion in $\alpha$-cells of pancreas, schematically presented in Figure~\ref{f.sch}, are briefly discussed, in accordance with the papers:~\cite{bb1,bb2,bb3}.

\subsection{The islets of Langerhans and the main hormones}
The secretion of insulin and glucagon occurs in the endocrine cells of the islets of Langerhans. These islets are located in the pancreas and constitute a small percentage of its total mass.  The pancreas mainly produces digestive enzymes secreted into the digestive tract, but it has also an endocrine function. the islets of Langerhans contain four principal cell types:  beta-cells producing insulin and amylin, which constitute 90\% of the islet, alfa-cells producing glucagon, gamma-cells producing pancreatic polypeptide, and delta-cells producing somatostatin. The islets are richly vascularized and hormones are directly secreted into bloodstream.\\
\newline
\textbf{Insulin}\\
The insulin molecule consists of two polypeptide chains: the A chain (21 - amino-acid) and B chain (30 residues) linked by two disulphide bridges.
Insulin exerts its main biological effects by binding to a cell-surface receptor.
Insulin acts in peripheral tissues through receptors located on the surface of muscle cells and adipocytes (fat cells). After insulin binds to the receptor, glucose channels are activated, allowing glucose to enter the cells. This mechanism leads to a decrease in blood glucose concentration. \\
\newline
\textbf{Glucagon}\\
Glucagon is a 29-amino-acid peptide that is secreted by alfa-cells from proglucnagon.
Its receptors are mainly found in the liver. 
Once glucagon is secreted into the circulation, it elicits its function intracellularly by
binding to its cell surface receptor G protein-coupled receptors (GPCRs). Glucagon evokes a signaling cascade causes the expression of gluconeogenic and glycogenolytic  process. Glucagon has the opposite effect to insulin. When blood sugar levels drop, glucagon is released into the bloodstream, signaling the liver to convert stored glycogen into glucose and  stimulates the process of gluconeogenesis. This process helps raise blood glucose levels, providing energy to the body, especially during periods of fasting or low carbohydrate intake. Glucagon works in opposition to insulin, which lowers blood sugar levels. It is essential in regulating energy balance and supporting metabolic functions.

\subsection{A pathway model of glucose-stimulated insulin secretion by $\beta$-cells}
\label{bi}
The primary stimulator for insulin secretion is the rising concentration of glucose in the blood. Therefore, its secretion is related to food intake. It also depends on changes the levels of amino acids and free fatty acids (FFA). Insulin secretion is modulated by the autonomic nervous system and influenced by incretins (GLP-1 glucagon-like peptide 1) which an enterohormones produced by intestinal cells.  

Insulin secretion is released in response to elevated blood glucose levels, such as after a meal.  Insulin is primarily secreted in a biphasic manner: a rapid release of pre-formed insulin and, after a short lag, a sustained release of newly synthesized insulin. The process begins when glucose enters the beta cells through glucose transporters, particularly GLUT2. Once inside, glucose undergoes glycolysis and subsequently enters the metabolic pathway of oxidative phosphorylation, leading to the production of ATP. As the ATP levels rise, they cause the closure of ATP-sensitive potassium channels in the cell membrane. This closure leads to depolarization of the beta cell membrane, triggering the opening of voltage-gated calcium channels. The influx of calcium ions into the beta cells is a key step; it stimulates the fusion of insulin-containing secretory granules with the plasma membrane. This fusion results in the exocytosis of insulin into the bloodstream. The secretion of insulin is also influenced by other factors, including hormones and nutrients, such as amino acids and fatty acids, which can further enhance or modulate the secretion response.

\subsection{A pathway model of glucose-stimulated glucagon secretion by $\alpha$-cells}
\label{bg}
The primary stimulus for glucagon release is a decrease in blood glucose levels. When blood glucose levels drop such as during fasting or prolonged periods without food intake or intense physical activity.  
Pancreatic $\alpha$-cells are equipped with a specific set of channels of Na+ and Ca2+, which  at low levels of glucose triggers Ca2+ signals and finally glucagon secretion. ATP-dependent K (K-ATP) channels plays a fundamental role in $\alpha$-cells, such as they do in $\beta$-cells, since they couple variations in extracellular glucose concentrations to changes in membrane potential and electrical activity.

Also, paracrine factors affect glucagon secretion. Insulin receptors are present on $\alpha$-cells. Additionally, insulin may indirectly suppress glucagon secretion through increasing
translocation of $\alpha$-cell GABA-A receptors (420). Inhibition of GABA receptors increases
glucagon secretion and GABA released from $\beta$-cells is postulated to mediate glucose-facilitated inhibition of glucagon secretion.

The pancreas is highly innervated by both the sympathetic (splanchnic) and parasympathetic
(vagus) nervous system. Centrally regulated glucagon secretion could be mediated via direct sympathetic innervation on the $\alpha$-cell, indirectly via the sympathetic tone and signaling through the hypothalamic-adrenal-pancreas signaling axis, and/or potential indirect parasympathetic signaling. Altogether, glucagon secretion is a complex process regulated by multiple interactions between glycemic, paracrine, endocrine, and neural factors.

In our current work, we focus on the primary stimulus of glucagon secretion -- changes in blood glucose levels. However, mentioned above, the more complex regulation mechanisms of glucagon secretion would be included in the next, future parts of the PN model.

\section{Models}
\label{models}
In this section, we present and discuss Petri net models of insulin secretion in pancreatic $\beta$-cells and glucagon secretion in pancreatic $\alpha$-cells. We also introduce a general model showing a joint operation of both individual models.

\subsection{Insulin secretion}
\label{mi}
%
Let us first discuss the model of insulin production, depicted on the left side of Figure~\ref{f.mi}, while its reachability graph is shown on the right hand side.\\
\begin{figure}[htb]
\begin{center}
\includegraphics[scale=0.33]{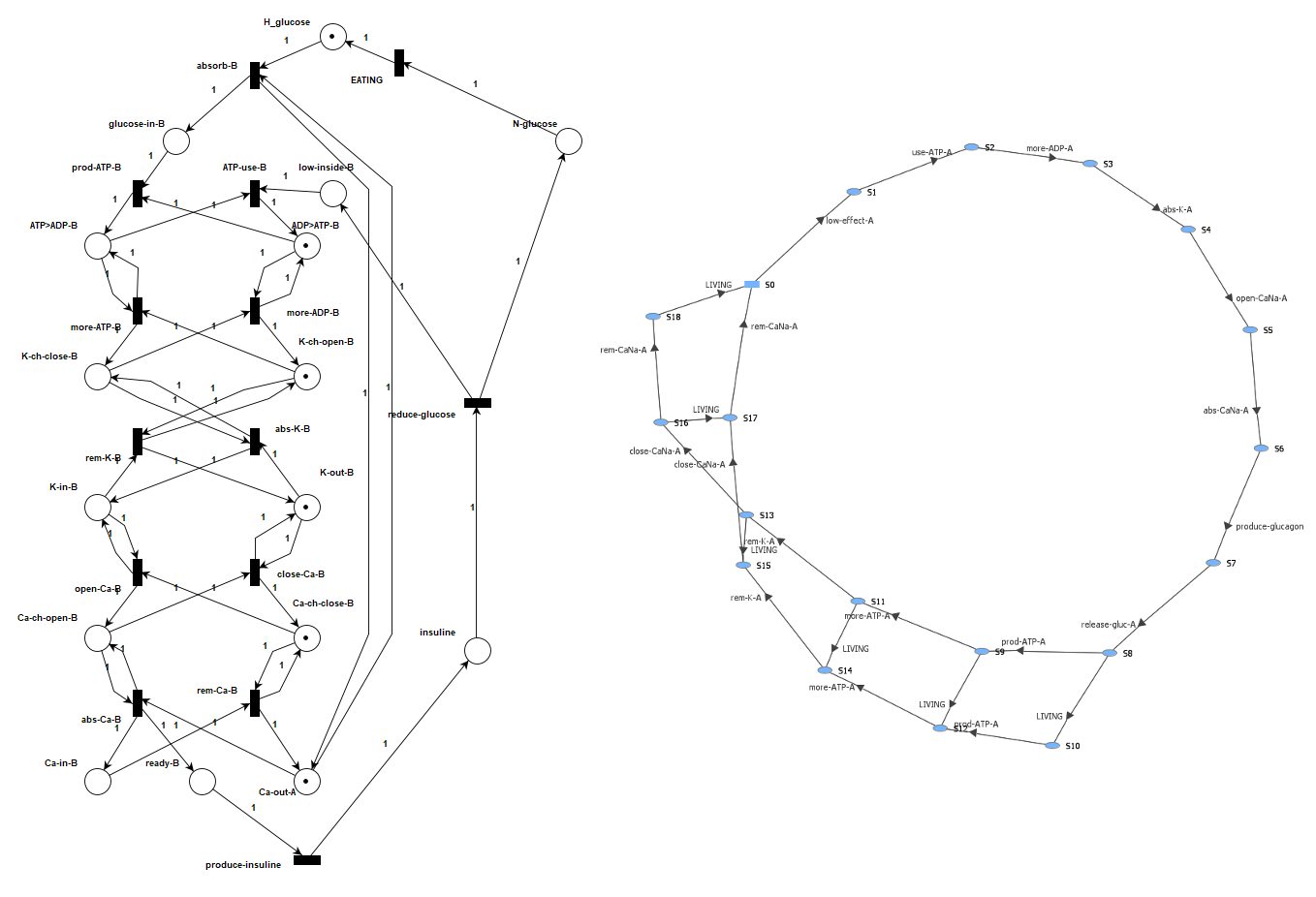}
\end{center}
\caption{Petri net modelling insulin production in pancreatic $\beta$-cells, together with its reachability graph (created with Pipe2~\cite{pipe2}).}
\label{f.mi}
\end{figure}

High glucose level is represented by the token located in \textbf{\textsf{H\_glucose}} place. 
Glucose is taken up by the $\beta$-cell of the pancreas, which is represented by the \textbf{\textsf{absorb-B}} transition. When present in the cell, which is modelled by the token in place \textbf{\textsf{glucose-in-B}}, it
is used in ATP production - represented by the \textbf{\textsf{prod-ATP-B}} transition.
The execution of this transition causes a token to appear in place \textbf{\textsf{ATP>ADP-B}}, which means that ATP/ADP ration
is in favour of ATP. In this situation, \textbf{\textsf{more-ATP-B}} transition becomes enabled. 
During execution, this transition transfers a token from place \textbf{\textsf{K-ch-open-B}} to place \textbf{\textsf{K-ch-close-B}} and it
corresponds to the situation when potassium channels become closed. 
When the channels are closed, transition \textbf{\textsf{abs-K-B}} becomes
enable -- it represents the increment of potassium in the cell by
moving a token from place \textbf{\textsf{K-out-B}} (representing potassium outside of the cell) to place \textbf{\textsf{K-in-B}} (representing potassium inside the cell).
A token in place \textbf{\textsf{K-in-B}} (corresponding to the higher amount of potassium in the cell) enables transition \textbf{\textsf{open-Ca-B}}.
This transition represents opening of the calcium channels by moving 
the token from \textbf{\textsf{Ca-ch-close-B}}, which represents closed channels, to place \textbf{\textsf{Ca-Ch-open-B}}, which represents open channels.
Then the transition \textbf{\textsf{abs-Ca-B}} is enabled, which models the process of increasing a level of calcium in the cell. This is done by moving a token from place \textbf{\textsf{Ca-out-B}} (calcium outside of the cell) 
to place \textbf{\textsf{Ca-in-B}} (calcium inside the cell)
and place \textbf{\textsf{ready-B}} (conditions are suitable to release 
insulin). 
A token in place \textbf{\textsf{ready-B}}, representing the suitable
conditions to secrete insulin, makes transition \textbf{\textsf{produce-insuline}} 
enabled. This transition corresponds to insulin secretion and produces a token to the place \textbf{\textsf{insuline}}.
Then transition \textbf{\textsf{reduce-glucose}} can be executed.
It represents a decrease in glucose levels and moves tokens to
places \textbf{\textsf{low-inside-B}} and \textbf{\textsf{N\_glucose}}
(symbolizes a normal level of glucose).

When glucose levels return to a normal (appropriate) value, a token appears in place \textbf{\textsf{low-inside-B}}. Then transition \textbf{\textsf{ATP-use-B}} is enabled, which represents the decrease in ATP level and the increase in ADP level.
This is modelled by relocating a token from place \textbf{\textsf{ATP>ADP-B}}
to \textbf{\textsf{ADP>ATP-B}}. 
The token in place \textbf{\textsf{ADP>ATP-B}} enables transition \textbf{\textsf{more-ADP-B}}, which moves the token from
 \textbf{\textsf{K-ch-close-B}} to place \textbf{\textsf{K-ch-open-B}}, which represents the opening of the potassium channels. 
When these channels are open, transition \textbf{\textsf{rem-K-B}} can
be executed (it represents the process of potassium leaving from the cell)
and it moves a token from place \textbf{\textsf{K-in-B}} to place the \textbf{\textsf{K-out-B}}.
When the potassium level in the cell decreases, which in the model is represented by a token in place \textbf{\textsf{K-out-B}} (i.e. ``potassium out''), transition \textbf{\textsf{close-Ca-B}} becomes enabled.
It represents closing of calcium channels and transfers a token from place \textbf{\textsf{Ca-open-B}} to place \textbf{\textsf{Ca-close-B}}.
This allows to execute transition \textbf{\textsf{rem-Ca-B}} -- 
corresponding to the process of decreasing of calcium level in the cell.
The transition transfers tokens from place \textbf{\textsf{Ca-in-B}} to place \textbf{\textsf{Ca-out-B}}. 
The token in place \textbf{\textsf{Ca-in-B}} means that the $\beta$-cell
goes back back completely to the state when the secretion of insulin is not possible.

After execution of transition \textbf{\textsf{EATING}} (representing ``eating'') the whole process starts all over again.

It is easy to observe that the net is bounded (as the reachability graph is finite). It is also reversible. 
Indeed, the parts of the model responsible for insulin production induced by high glucose levels,  i.e.:
\begin{itemize}
\item [(I)]\textbf{\textsf{prod-ATP-B}}$\Rightarrow$
\textbf{\textsf{ATP>ADP-B}} $\Rightarrow$ 
\textbf{\textsf{more-ATP-B}}$\Rightarrow$
\textbf{\textsf{K-ch-close-B}}$\Rightarrow$
\textbf{\textsf{abs-K-B}}$\Rightarrow$
\item []\textbf{\textsf{K-in-B}}$\Rightarrow$
\textbf{\textsf{open-Ca-B}}$\Rightarrow$
\textbf{\textsf{Ca-ch-open-B}}$\Rightarrow$
\textbf{\textsf{abs-Ca-B}}$\Rightarrow$
\textbf{\textsf{Ca-in-B}}, \textbf{\textsf{ready-B}},
\end{itemize}
and for the return of the $\beta$-cell to the initial situation, i.e.
\begin{itemize}
\item [(II)]\textbf{\textsf{ATP-use-B}}$\Rightarrow$
\textbf{\textsf{ATP<ADP-B}} $\Rightarrow$ 
\textbf{\textsf{more-ADP-B}}$\Rightarrow$
\textbf{\textsf{K-ch-open-B}}$\Rightarrow$
\textbf{\textsf{rem-K-B}}$\Rightarrow$
\item []\textbf{\textsf{K-out-B}}$\Rightarrow$
\textbf{\textsf{close-Ca-B}}$\Rightarrow$
\textbf{\textsf{Ca-ch-close-B}}$\Rightarrow$
\textbf{\textsf{rem-Ca-B}}$\Rightarrow$
\textbf{\textsf{Ca-out-B}}, 
\end{itemize}
execute alternately, which is observable in
the reachability graph (Figure~\ref{f.mi}).

Moreover, switching from the second path (returning
to the initial state) to the first one 
(production of insulin) is possible only after execution 
of transition \textbf{\textsf{EATING}}. It corresponds
to the real, biological behavior, that after insulin
is released (and the level of glucose decreases),
secretion of insulin stops and the cell returns to 
its initial, nonactive state. And after
eating, the increasment in  glucose level
initializes processes resulting in insulin secretion.

\subsection{Glucagon secretion}
\label{mg}

\begin{figure}[htb]
\begin{center}
\includegraphics[scale=1.6]{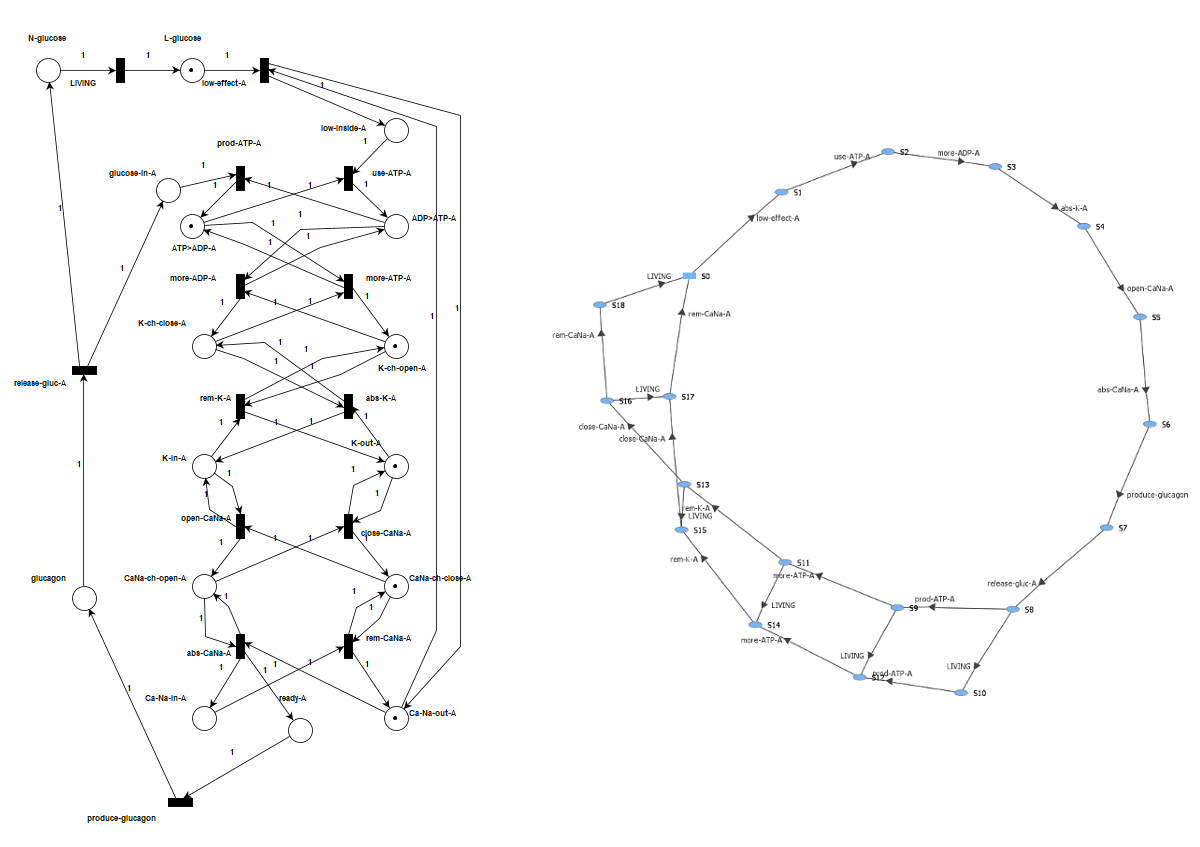}
\end{center}
\caption{Petri net modelling glucagon production in pancreatic $\alpha$-cells, together with its reachability graph (created with Pipe2~\cite{pipe2}).}
\label{f.mg}
\end{figure}

Let us now focus on the model of glucose-dependent glucagon secretion, depicted on the left side of Figure~\ref{f.mg}, while its reachability graph is shown on the right hand side.

In the case of the glucagon production, most of the model looks similar to the insulin production one, except that the names of the elements ends with ``\textbf{\textsf{A}}''. 
The main difference is, however, that in this case the potassium channels close when the token is located in place \textbf{\textsf{ADP>ATP-A}}, 
and the channels open when the token is located in place \textbf{\textsf{ATP>ADP-A}}. 
Transition \textbf{\textsf{more-ATP-A}} is responsible for opening
the channels, and transition \textbf{\textsf{more-ADP-A}} for closing.
Furthermore, since in $\alpha$-cells not only calcium channels must be open, but also sodium ones, the model includes places \textbf{\textsf{CaNa-ch-open-A}} and \textbf{\textsf{CaNa-ch-close-A}}, 
analogous to places \textbf{\textsf{Ca-open-B}} and \textbf{\textsf{Ca-close-B}} from the insulin production model.

The reachability graph presents the dynamic of the PN model. Same as in the case of
 of insulin production by $\beta$-cells, the model of $\alpha$-cells
production of glucagon is also bounded and reversible. 
The two processes are performed alternately. 
First one represents the secretion of glucagon and consists of:
\begin{itemize}
\item [(III)]\textbf{\textsf{use-ATP-A}}$\Rightarrow$
\textbf{\textsf{ADP>ATP-A}} $\Rightarrow$ 
\textbf{\textsf{more-ADP-A}}$\Rightarrow$
\textbf{\textsf{K-ch-close-A}}$\Rightarrow$
\textbf{\textsf{abs-K-A}}$\Rightarrow$
\item []\textbf{\textsf{K-in-A}}$\Rightarrow$
\textbf{\textsf{open-CaNa-A}}$\Rightarrow$
\textbf{\textsf{CaNa-ch-open-A}}$\Rightarrow$
\textbf{\textsf{abs-CaNa-A}}$\Rightarrow$
\textbf{\textsf{Ca-in-A}}, \textbf{\textsf{ready-A}}.
\end{itemize}
The second corresponds to the process of returning of the
$\alpha$-cell to its initial state and consists of:
\begin{itemize}
\item [(IV)] \textbf{\textsf{prod-ATP-A}}$\Rightarrow$
\textbf{\textsf{ATP>ADP-A}} $\Rightarrow$ 
\textbf{\textsf{more-ATP-A}}$\Rightarrow$
\textbf{\textsf{K-ch-open-A}}$\Rightarrow$
\textbf{\textsf{rem-K-A}}$\Rightarrow$
\item []\textbf{\textsf{K-out-A}}$\Rightarrow$
\textbf{\textsf{close-CaNa-A}}$\Rightarrow$
\textbf{\textsf{CaNa-ch-close-A}}$\Rightarrow$
\textbf{\textsf{rem-CaNa-A}}$\Rightarrow$
\textbf{\textsf{CaNa-out-A}}.
\end{itemize}
It is easy to observe in the graph, that to 
start the process resulting in glucagon secretion, it is
necessary to execute the \textbf{\textsf{LIVING}} transition.
This transition represents the normal, biological consumption 
of glucose by cells to perform their living functions.

\subsection{Combined model}
\label{mc}

\begin{figure}[htb]
\begin{center}
\includegraphics[scale=0.35]{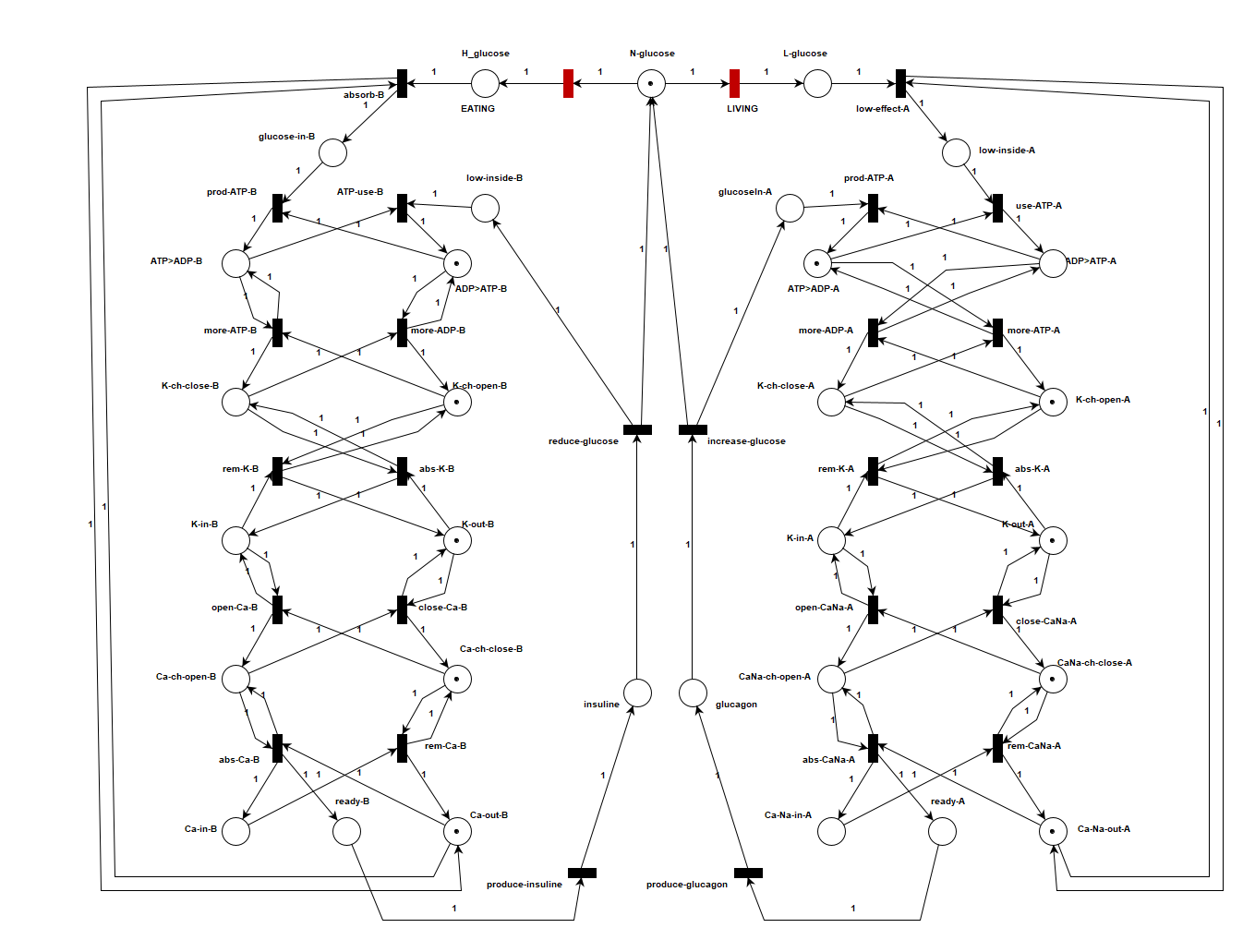}
\end{center}
\caption{A Petri net modelling glucagon and insulin production in pancreatic $\alpha$- and $\beta$-cells (created with Pipe2~\cite{pipe2}).}
\label{f.mc}
\end{figure}

In this section, we analyse the combined model, depicted in Figure~\ref{f.mc}, in which both insulin and glucagon can be secreted, depending on the current blood glucose level. 

As before, places \textbf{\textsf{H\_glucose}}, \textbf{\textsf{N\_glucose}}, \textbf{\textsf{L\_glucose}} represent high, normal, and low blood glucose levels, respectively. Let us note that if in the initial
state, i.e. at a normal blood glucose level, transition
\textbf{\textsf{EATING}} executes, then the token moves from place \textbf{\textsf{N\_glucose}} to the place \textbf{\textsf{H\_glucose}}, makes transition \textbf{\textsf{absorb-B}} fireable
and enables execution of transition the left part of the net, responsible for insulin secretion. 
On the other hand, when transition \textbf{\textsf{LEAVING}} is
executed, then the token goes to place \textbf{\textsf{L\_glucose}}, and the right side of the net, modelling the secretion of glucagon, becomes
active (transition \textbf{\textsf{low-effect-A}} becomes enabled). 
As one might notice, the left and right parts of the net are constructed from the previously introduced individual models for insulin and glucagon
secretion, respectively.
 
\begin{figure}[htb]
\begin{center}
\includegraphics[scale=0.47]{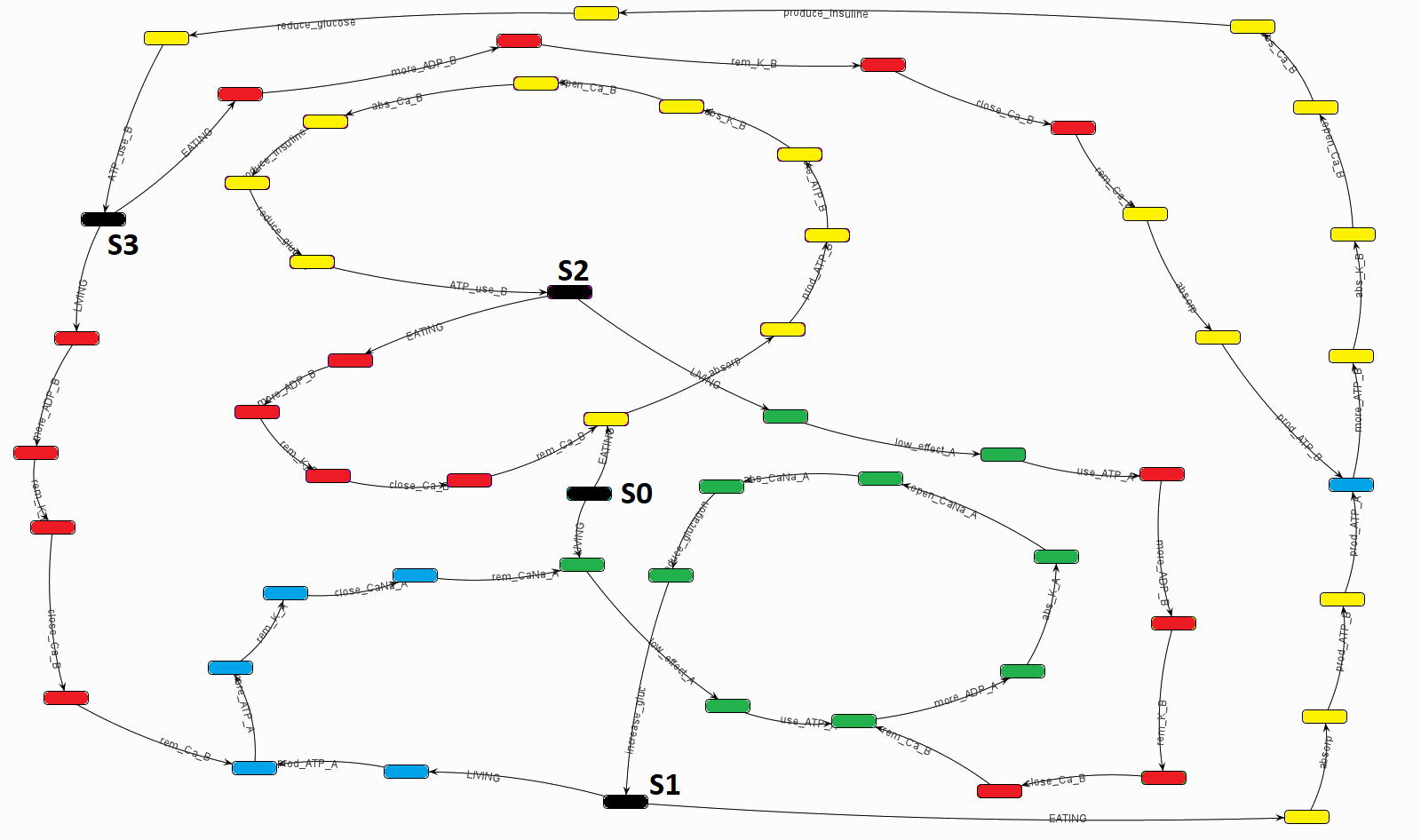}
\end{center}
\caption{A stubborn reduced reachability graph obtained from the Petri net modelling glucagon and insulin production, depicted in Figure~\ref{f.mc} (created with Snoopy/Charlie~\cite{snoopy, charlie}). Four states 
are distinguished, denoted by \textbf{\textsf{S0}}, \textbf{\textsf{S1}}, \textbf{\textsf{S2}} and \textbf{\textsf{S3}}.}
\label{f.g.mc}
\end{figure}

As known, the best way to analyse the behaviour of a Petri net, is by
examining its reachability graph. 
In this case however, even though the net is bounded, we do not show its reachability graph here, because it contains 192 states with 418 arc, and is too large to be displayed in a transparent way.
Instead, we will settle for analysis of stubborn reduced reachability graph presented in Figure~\ref{f.g.mc}.
\newpage
In the graph four states have been distinguished: \textbf{\textsf{S0}} which corresponds to
the initial marking, \textbf{\textsf{S1}}, \textbf{\textsf{S2}} and \textbf{\textsf{S3}}. In each of these states
transitions \textbf{\textsf{EATING}} and \textbf{\textsf{LIVING}}
are enabled. These states are the vertices in which the paths in the graph intersect,
and the execution of one of the two mentioned transitions
determines the behaviour of the model. The paths between the highlighted states
correspond to the main processes described in previous sections
for the separate models: 
secretion of glucagon by $\alpha$-cells, returning to the inactive state of 
$\alpha$-cells, the secretion of insulin by $\beta$-cells and returning 
to the inactive state of $\beta$-cells.
The path from \textbf{\textsf{S0}} (the initial marking) to \textbf{\textsf{S1}},
which starts with transition \textbf{\textsf{LIVING}},
contains transitions related to the secretion of glucagon (states
marked in green). On the other hand, when at \textbf{\textsf{S0}} transition
\textbf{\textsf{EATING}} is executed, the state \textbf{\textsf{S2}} is reachable and the path between \textbf{\textsf{S0}} and \textbf{\textsf{S2}}  contains transitions representing the secretion of 
insulin (marked in yellow). 
Then, when at state \textbf{\textsf{S1}} transition
\textbf{\textsf{LIVING}} fires, the path leads back to state \textbf{\textsf{S1}}
and contains transitions related to returning to the inactive state
by $\alpha$-cells (marked in blue) and the secretion of glucagon.
Notice that, starting for the initial marking it would be
the second execution of transition \textbf{\textsf{LIVING}}.
Similarly, if at state \textbf{\textsf{S2}} transition \textbf{\textsf{EATING}}
is executed (starting for the initial marking it is
the second execution of that transition), then the path
in the graph leads back to state \textbf{\textsf{S1}} and contains 
transitions representing the process of  
returning to the inactive state
by $\beta$-cells (marked in red) and the secretion of insulin.
If in \textbf{\textsf{S1}} transition \textbf{\textsf{EATING}} fires,
state \textbf{\textsf{S3}} is reached and the path corresponding to the secretion of insulin.
From state \textbf{\textsf{S2}} state \textbf{\textsf{S1}} can be reached by  firing of transition
\textbf{\textsf{LIVING}} and the path containing elements 
related to the process of returning to the inactive state
by $\beta$-cells and the secretion of glucagon.
Note that, starting from the initial marking,
the path from \textbf{\textsf{S0}}, then \textbf{\textsf{S2}} to \textbf{\textsf{S1}} includes 
transitions \textbf{\textsf{EATING}}
and \textbf{\textsf{living}}, each once.
State \textbf{\textsf{S1}} can be reached from state \textbf{\textsf{S3}} 
by firing of transition \textbf{\textsf{LIVING}}
and the path corresponds to the
process of returning to the inactive state 
by $\alpha$-cells and $\beta$-cells, and the secretion of glucagon.
To reach \textbf{\textsf{S1}} through \textbf{\textsf{S3}} from \textbf{\textsf{S0}} transition \textbf{\textsf{EATING}}
has to be executed once and transition \textbf{\textsf{LIVING}} twice.
When at \textbf{\textsf{S3}} transition \textbf{\textsf{EATING}} fires,
the path leads back to \textbf{\textsf{S3}} and it is related to the process of returning to the inactive state
by $\beta$-cells and the secretion of insulin.
To conclude this part of the analysis, one can notice
that every time transition \textbf{\textsf{EATING}} fires,
the part of the model corresponding to the secretion of insulin is active. 
Eventually, before the next secretion, $\beta$-cells have to return 
to their initial state. It happens when transition \textbf{\textsf{EATING}}
has been previously executed. Similarly, after execution
of transition \textbf{\textsf{LIVING}}, the secretion of glucagon takes place. 
If that transition has been executed previously, 
$\alpha$-cells have to go back to their inactive state
before the next secretion. This behaviour is
desirable and consistent with biological processes,
which is our most important goal.

One might notice, that in the reachability graph presented
in Figure~\ref{f.g.mc} the model does not go back
to the initial state. This is effect of the stubborn reduction.
The model is live, reversible and it is covered by $T$-invariants.
All those properties are important and desirable in biological 
models. 

\begin{figure}[htb]
\begin{center}
\includegraphics[scale=0.8]{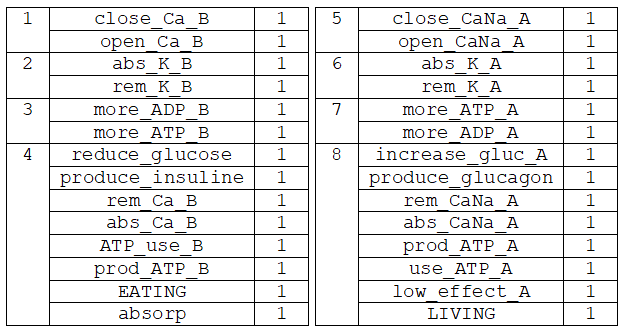}
\end{center}
\caption{Two semipositive $T$-invariants of the Petri net depicted in Figure~\ref{f.mc}.}
\label{f.inv}
\end{figure}

In Figure~\ref{f.inv} we present all the semipositive $T$-invariants of the Petri net depicted in Figure~\ref{f.mc}. Every transition is included in a 
$T$-invariant. Moreover, $T$-invariants
$1$, $2$, $3$ and $4$ contain every transition from the insulin part 
of the model, $T$-invariants $5$, $6$, $7$ and $8$ -- every transition
from the glucagon part of the model. Like mentioned in 
Section~\ref{s.prel.pn}, the execution of transitions from $T$-invariant
does not result in the change of the marking. Hence, by starting
from the initial marking and executing transitions from $T$-invariants:
$1$, $2$, $3$ and $4$ the initial marking would be reached. 
The same from transitions for $T$-invariants: $5$, $6$, $7$ and $8$.
It shows that the model presents a cyclic behaviour.

To summarize, the models precisely reflects the biological processes
of the secretion of insulin, glucagon, their relations and cyclicality.
It is important to notice, that when the most of transitions precisely
model the exact biological process, some transitions symbolize
more complex effects. This applies to transitions: \textbf{\textsf{EATING}},  \textbf{\textsf{LIVING}}, \textbf{\textsf{reduce-glucose}} and
\textbf{\textsf{increase-glucose}}.
The first one represents all
processes, which can elevate the glucose level. The most
common one is, indeed, eating, which provides carbohydrates.
The second one corresponds to processes of reducing glucose level.
The most important one is usage of glucose by cells to perform their
living functions. In both cases other phenomena may affect
the level of glucose.
The process of increasing the glucose level, modelled by
transition \textsf{reduce-glucose}, has the same effect 
like in the model, but in organisms it consists of many 
other processes, like absorption of glucose by cells or storing
of glucose in the liver or the fat tissue. All those complex effects
are represented by only one transition in the model.
The same occurs for transition \textsf{increase-glucose}.
This process is complex, and involves releasing of glucose
stored in different parts of the body. It the model
it is represented by a single transition.
A part of this complex process is modelled in our previous paper (\cite{BG1}), and ultimately we intend to redesign the model in such a way that individual transitions are replaced by distinct processes represented in a form of a Petri net models.

\section{Boolean network}
\label{bn}

As could be seen in the previous sections, describing biological foundations and models, glucose-stimulated insulin secretion by $\beta$-cells as well as glucose-stimulated glucagon secretion by $\alpha$-cells on the pancreas take the form of pathways, i.e. sequences of processes undergone by a specific compounds.
Note that the presented Petri net models operate largely on the binary principles, where a token in a~given place symbolizes the occurrence of a certain phenomenon, while the absence of a token means that the phenomenon does not occur.
In fact, all presented PN models are \textit{safe} (or $1$-safe), which means that the number of tokens in any place at any reachable marking does not exceed $1$.

On the other hand, one of the commonly used models to study complex dynamic behaviour of a~biological systems, which perfectly model biological pathways, are Boolean networks. In many scientific papers (among others~\cite{pnbn0,pnbn2,pnbn3,pnbn1}) one can find Petri nets and Boolean networks constituting a common area of interest. However, due to the general specificity of both of these models, transformations leading from Boolean networks to Petri nets are more common.

\begin{figure}[htb]
\begin{center}
\includegraphics[scale=0.93]{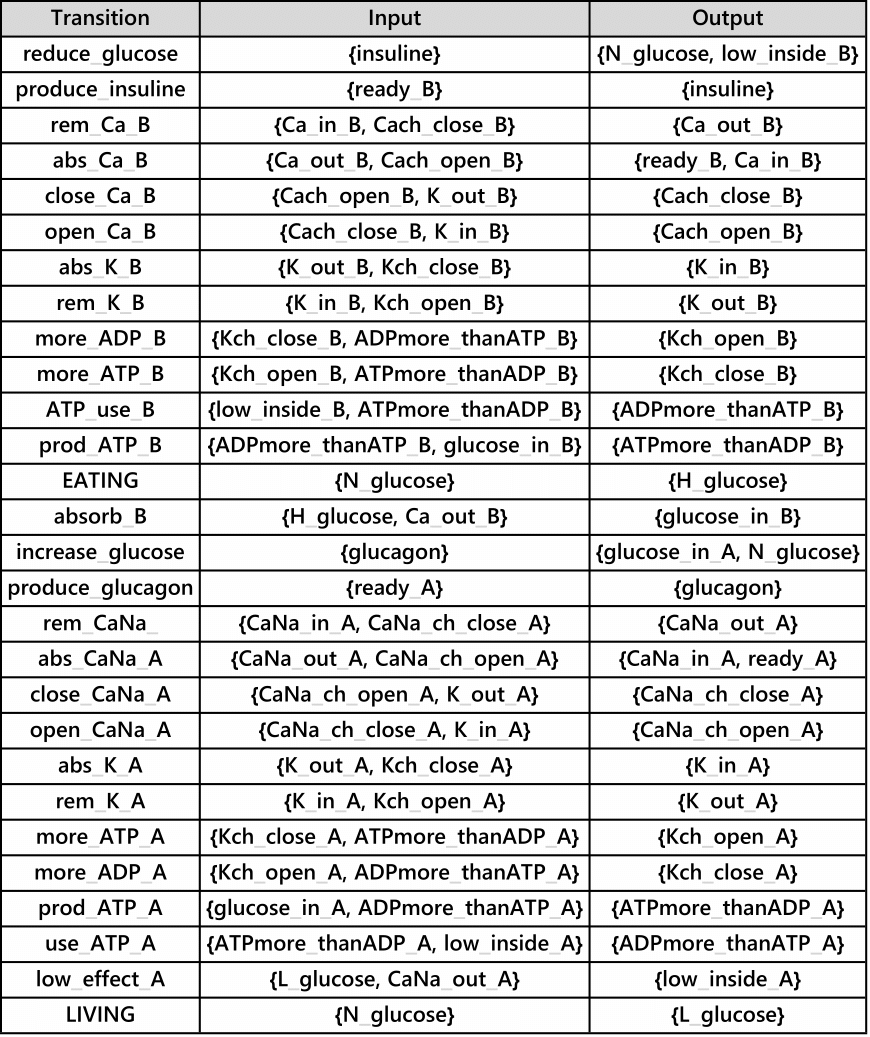}
\end{center}
\caption{The model table of the Petri net shown in Figure~\ref{f.mc}.}
\label{f.mtbn1}
\end{figure}

\begin{figure}[htb]
\begin{center}
\includegraphics[scale=1.15]{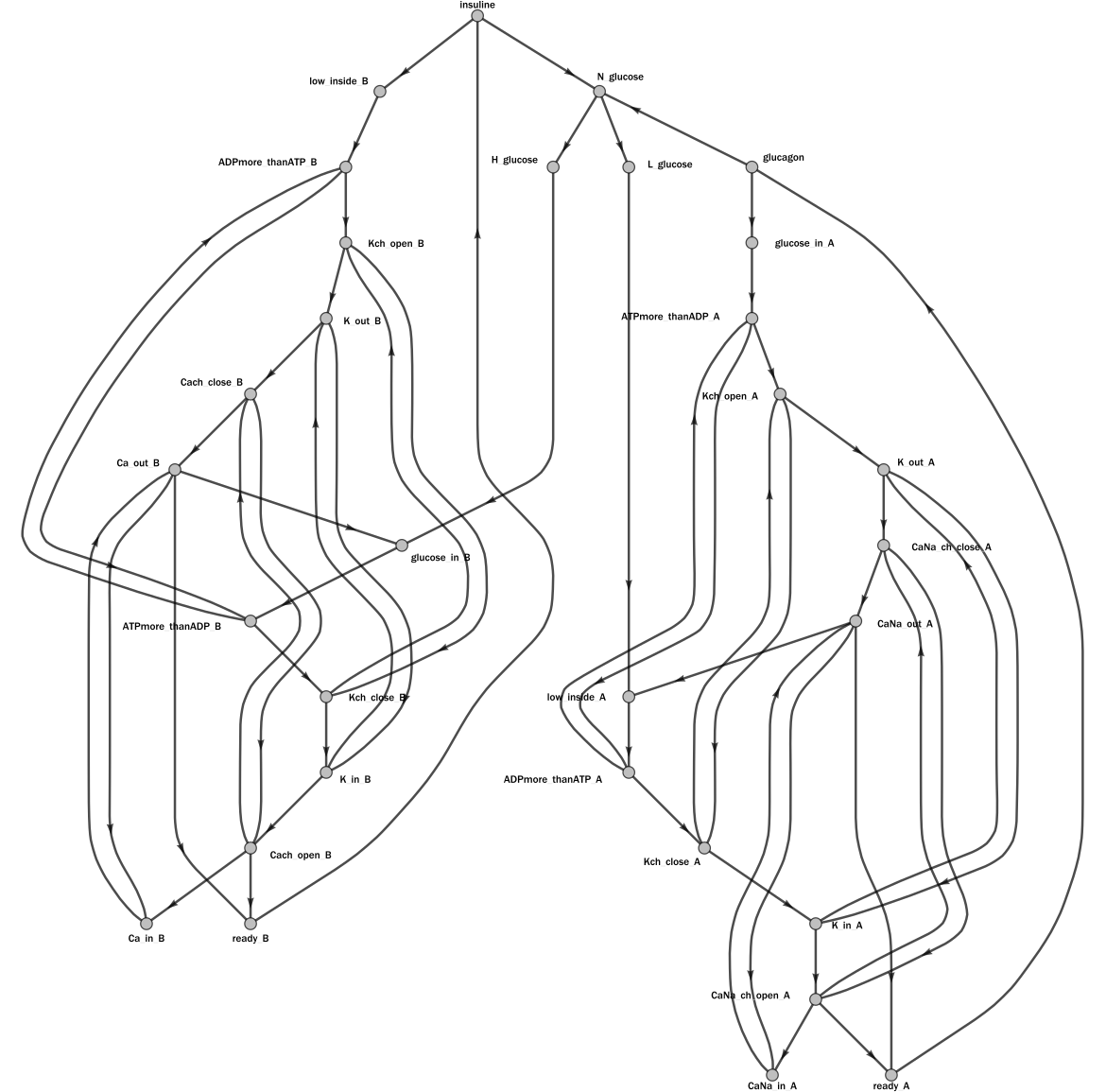}
\end{center}
\caption{The interaction graph of the Petri net shown in Figure~\ref{f.mc}.}
\label{f.igbn1}
\end{figure}

In this paper we decided to transform the combined Petri net model presented in Section~\ref{mg} into a~Boolean network. The transformation rules are rather intuitive and are based on the possibility of a~token moving (by execution 
of a single transition) from one place to another. Describing pre-sets and post-sets of every transition, we obtain the net interaction characteristics. Due to the fact that self-loops from the original Petri net have no impact on the flow, we decided not to include them in the model, therefore the model table (Figure~\ref{f.mtbn1}) the interaction graph (Figure~\ref{f.igbn1}) do not include them.

Let $N=(P,T,F,M_0)$ be a Petri net depicted in Figure~\ref{f.mc}. In order to transform $N$ into a Boolean net, we use the rules described below.
 
Let $p\in P$ be a place, and let $\{t_1,t_2,\ldots,t_k\}={^\bullet}p$ be a pre-set of $p$, i.e. the set of transitions for which $p$ is an output place. For $t_i\in {^\bullet}p$ let $\{p_{i_1},p_{i_2},\ldots p_{i_{l_i}}\}$ be a set of all places being an entry to $t_i$. Then in the Bollean network we need to introduce the following formula:
$$p=(p_{1_1}\land p_{1_2}\land\ldots\land p_{1_{l_1}})\lor (p_{2_1}\land p_{2_2}\land\ldots\land p_{2_{l_2}})\lor\ldots\lor (p_{k_1}\land p_{k_2}\land\ldots\land p_{1_{l_k}}).$$

\begin{figure}[htb]
\begin{center}
\includegraphics[scale=1.1]{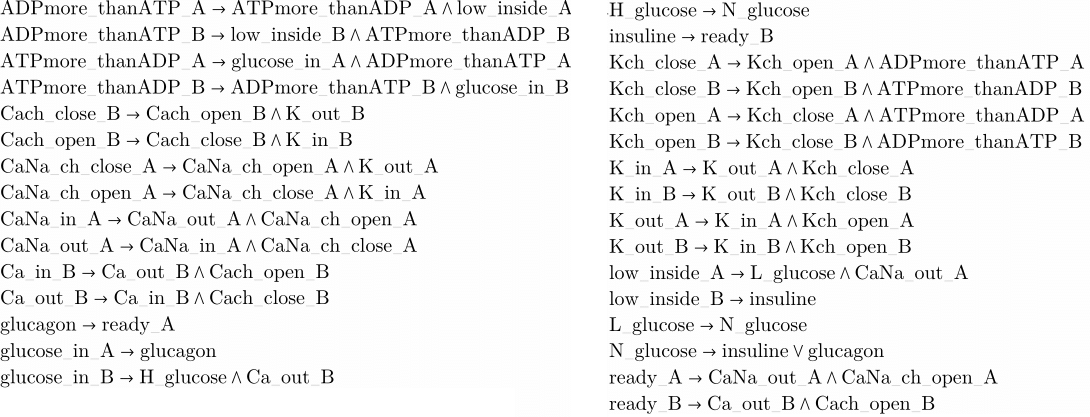}
\end{center}
\caption{The Boolean network based on the Petri net shown in Figure~\ref{f.mc}.}
\label{f.bn1}
\end{figure}

By performing the transformation based on the above rule for all locations in the model, we obtain the  Boolean network presented in Figure~\ref{f.bn1}. 

\begin{figure}[htb]
\begin{center}
\includegraphics[scale=0.25]{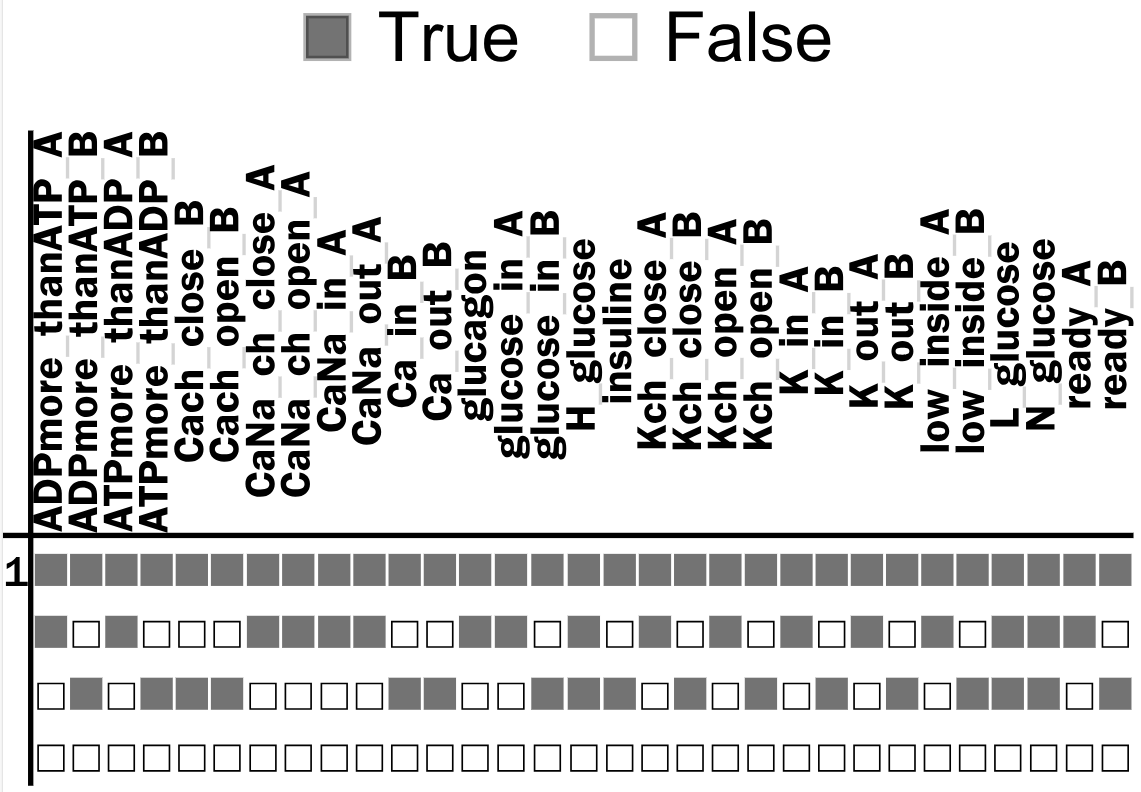}
\end{center}
\caption{The set of stable states of a Boolean network presented in Figure~\ref{f.bn1}.}
\label{f.ssbn1}
\end{figure}

As described, the Petri Net model of insulin secretion and glucagon secretion in 
the pancreas can be easily transformed into a Boolean network and any tools that allow analysis of Boolean networks can be used for its further analysis. This approach provides the future opportunity to analyse both models simultaneously or use them alternately depending on current needs.
We can, for instance, compute all stable states for our Boolean network (depicted in Figure~\ref{f.ssbn1}).
In calculated stable states of the model, either all variables are assigned $0$s, or all of them are assigned $1$s. Another possibilities are (1) all variables associated with the left part of the original net (corresponding to the insulin secretion), together with those connected to glucose levels, are assigned $1$s, and the remaining ones (corresponding to the glucagon secretion) are assigned $0$s or vice versa, (2) variables associated with glucagon secretion (with those connected to glucose levels) are assigned  $1$s, and those connected to the insulin secretion are assigned $0$s. This reflects well the
dynamic aspects of the model.

\section{Conclusions and Future Work}

In the paper we took another step towards modelling all processes responsible for the regulation of glycemia in humans. We presented Petri net models of insulin secretion in $\beta$-cells of the pancreas and glucagon secretion in $\alpha$-cells of the pancreas, as well as a comprehensive model taking into account both processes. We also analysed the dynamics and properties of the three models.
Additionally, we~presented a transformation of our comprehensive model to Boolean networks.
\newpage
Our analysis shows that the PN model is able to reproduce the biological processes underling it. 
Like in the real body, when the level of glucose is high, insulin is secreted, and when the level of glucose is low, glucagon is produced. This mimic the real, biological behaviour, which was our goal. 
One~might notice the one-dimensionality of the presented model in relation to the ``signalling pathway'' of $\alpha$- and $\beta$- cells. 
Indeed, since this is our first approach, it does not take into account, among other things, the interaction between insulin secretion and glucagon secretion and vice versa: increased glucagon secretion under the influence of increased insulin concentration (postprandial). Other factors modelling the secretion of both hormones, such as incretins (endocrine hormones), were not taken into consideration, too. 
Our future plans include modelling these interactions as well. As a result of our work, we hope to obtain a~comprehensive model of glycemia regulation in humans.

On the other hand we would like to develop a solution allowing the simultaneous or alternating use of various mathematical models, such as Petri nets or Boolean networks, enabling the most accurate, but also as easy as possible, analysis of biological systems.

\newpage

\end{document}